\documentclass{article}

\usepackage{amssymb,amsmath,amsthm,bbm}
\usepackage{verbatim,float,url,dsfont}
\usepackage{graphicx,subcaption,psfrag}
\usepackage{algorithm}
\usepackage[noend]{algorithmic}
\usepackage{mathtools,enumitem}
\usepackage{multirow}
\usepackage{ragged2e}
\usepackage{xr-hyper}
\usepackage{caption}
\usepackage{array}

\usepackage[colorlinks=true,citecolor=blue,urlcolor=blue,linkcolor=blue]{hyperref}
\usepackage[margin=1in]{geometry}

\usepackage[utf8]{inputenc} % allow utf-8 input
\usepackage[T1]{fontenc}    % use 8-bit T1 fonts
\usepackage{booktabs}       % professional-quality tables
\usepackage{nicefrac}         % compact symbols for 1/2, etc.
\usepackage{microtype}      % microtypography

\usepackage[dvipsnames]{xcolor} % Colored text
\usepackage{pifont} % For symbols like checkmarks, see http://ctan.org/pkg/pifont
\usepackage{etoc} % For local table of contents within sections "\localtableofcontents"
\usepackage{tikz} % for diagrams
\usepackage{pgfplots}  % for diagrams
\pgfplotsset{compat=1.16}  % for diagrams 

\ifdefined\TimesFont 
\usepackage{times} % use times font
\fi

\ifdefined\ParSkip 
\usepackage{parskip} % use par skip
\fi

\theoremstyle{definition}

\newtheorem*{assumption*}{\assumptionnumber}
\providecommand{\assumptionnumber}{}


\makeatletter
\newcommand*\rel@kern[1]{\kern#1\dimexpr\macc@kerna}
\newcommand*\widebar[1]{%
  \begingroup
  \def\mathaccent##1##2{%
    \rel@kern{0.8}%
    \overline{\rel@kern{-0.8}\macc@nucleus\rel@kern{0.2}}%
    \rel@kern{-0.2}%
  }%
  \macc@depth\@ne
  \let\math@bgroup\@empty \let\math@egroup\macc@set@skewchar
  \mathsurround\z@ \frozen@everymath{\mathgroup\macc@group\relax}%
  \macc@set@skewchar\relax
  \let\mathaccentV\macc@nested@a
  \macc@nested@a\relax111{#1}%
  \endgroup
}
\makeatother

\usepackage{tikz}

\usepackage[round]{natbib}

\title{A Collectivist, Economic Perspective on AI}    
\author{
Michael I.\ Jordan} 
\date{Inria Paris, and the
University of California, Berkeley} 

\begin{document}
\maketitle

\begin{abstract}
Information technology is in the midst of a revolution in which omnipresent data 
collection and machine learning are impacting the human world as never before.  
The word ``intelligence'' is being used as a North Star for the development of 
this technology, with human cognition viewed as a baseline.  This view neglects 
the fact that humans are social animals and that much of our intelligence is 
social and cultural in origin.  Moreover, failing to properly situate aspects
of intelligence at the social level contributes to the treatment of the societal
consequences of technology as an afterthought.  The path forward is not merely 
more data and compute, and not merely more attention paid to cognitive or symbolic 
representations, but a thorough blending of economic and social concepts with 
computational and inferential concepts at the level of algorithm design.
\end{abstract}

\section{Introduction}

The current dialogue on artificial intelligence (AI)---in the media, in academia, 
in industry, in government commissions, and around dinner tables---often seems 
untethered to reality.  The discussion tends to focus on whether one should be 
on the side of hype or hysteria---AI will either solve humanity's most pressing 
problems and usher in a new era of plenty or it will destroy or enslave the human 
species.  Whereas previous eras of rapid technological development were accompanied 
by such dialogue, the current extreme nature of the hype and hysteria seems unprecedented.

The phrase ``AI'' arose in the 1950s, and while the phrase was provocative and exciting, 
the action over the ensuing decades in computer science was elsewhere---in the development 
of hardware, languages, networks, search engines, human-computer interaction, and 
eventually data collection and machine learning.  The phrase ``machine learning'' 
(ML) was coined by an AI researcher~\citep{Samuel}, but it was eventually adopted 
by researchers in many other fields, including operations research, control theory, 
and statistics, who brought with them a range of experience and applications in 
engineering and science.  Machine learning thus served as an intellectual bridge 
for a data-intensive era---catalyzing the formation of cross-disciplinary connections 
and (critically) connecting mathematically inclined researchers from various backgrounds 
with the computer scientists who were building computing systems and networks of 
ever-increasing scale.

And then came large-language models (LLMs).  The ideas and architectures underlying 
LLMs were fully in the ML tradition---using gradient-based methods to adjust parameters 
in large-scale predictive systems---with the key novelty that the data was human 
language, in truly massive quantities. The output of LLMs was (strikingly) fluent 
language, giving the appearance of a human-like entity.  This triggered the return 
to prominence of the phrase ``AI.''

But whereas an LLM may appear to be a single ``entity'' that is human-like, it is 
equally well understood as a ``collectivist'' artifact.  Indeed, in interacting with 
an LLM, one is interacting implicitly with a vast number of humans who have contributed 
micro-level data, opinions, linguistic constructions, and creative works to the 
LLM via the Internet and other media.  When these human contributions agree in 
various ways, the LLM is able to promote that agreement into abstractions that are 
useful and that strengthen the illusion of personhood. But, while the analogy of 
an LLM to a person seems irresistible, an analogy of an LLM to a culture is equally 
valid.  Cultures are repositories of narratives, opinions, and abstractions.  
Cultures have personalities.

Moreover, if we wish to view LLMs as exhibiting human-like intelligence, 
and in particular as possessing real-world problem-solving skills, then we need 
to remember that human intelligence is in part social in nature, and that the 
success of individual efforts is often best measured in the overall social context 
in which they are embedded (even from the point of view of the individual).  
Accordingly, let us lift our eyes from the LLM to consider the ecosystems in 
which LLM algorithms are being embedded.  Consider in particular the planetary-scale 
networks that have emerged in our era to solve problems in domains such as 
commerce, healthcare, transportation, logistics, education and entertainment.  
These networks involve vast numbers of heterogeneous participants, some of 
whom are human and some non-human.  The participants are linked by flows of 
data that increasingly allow them to learn from each other.  Learning may
involve cooperation, competition, association or collusion.  Participants 
may share some of their resources, including global resources, but they may 
also want to hold some resources locally.  This will occur for many reasons, 
but most significantly because participants will have a desire to obtain value 
and competitive advantage from their particular knowledge, data, or creative output.

Thus, overall, as I will argue, an appropriate metaphor for emerging AI 
systems is closer to that of a \emph{market} than a search engine, a chatbot, 
or a personal secretary.  (The latter are mere roles in the overall market).  
This economic perspective allows a balanced consideration of both the producer 
role and the consumer role of the humans participating in the system. 
The consideration of the producer role has lagged that of the consumer
role, and people are increasingly asking what benefits accrue to them when 
their creative activity is used in the training of LLMs.  Such questions are 
not new---they also arose in the era of the search engine---but in that case,
individuals obtained clear value both as producers and consumers.  In their 
role as producers, they obtained visibility and traffic, and in their roles 
as consumers they obtained access to information, knowledge, and services.  
These benefits were part of an implicit social and economic contract in 
which data on the Internet was treated as free for the taking, and in 
exchange services were provided freely.  

That same contract is being offered (implicitly) by the companies developing LLMs.
But the scope of LLMs and generative AI goes far beyond search engines---the
goal is no longer to merely provide links that guide users to useful websites, 
but to aggregate and transform their input data so as to engage in sustained
dialog and creative activities.  A consequence is that the LLM becomes the 
endpoint rather than an intermediary, and the benefit of visibility and traffic 
for producers begins to wither---it is no longer part of the implicit contract.  

What will these new markets powered by data and machine learning look like?
As in the case of historical markets, bottom-up self-organization will be the 
dominant paradigm for growth of learning-catalyzed markets.  But such growth 
need not be uncontrolled or outside of our comprehension. 

The essential point is that these new markets are arising not because of 
a deep scientific understanding of the nature of human intelligence, but 
\emph{rather because of the flowering of the concept of an algorithm---one  
of one of the key achievements of the 20th century}.  Algorithms, as instantiated 
in hardware, software and mathematical models, are the way that information 
technology systems are developed.  We can gain leverage over emerging 
systems by considering more deeply the source of the algorithmic ideas 
that are driving the technology, asking about the extent to which they 
connect to societal-level goals, and expanding the space of algorithms if
they prove deficient.

In considering the nature and source of algorithmic ideas, it is useful to 
start with the phrase ``computational thinking.''  This phrase aims to 
capture the idea that the algorithmic concepts developed in computer 
science---such as modularity, abstraction, and scaling---have broad 
applicability to problem-solving activities throughout science and 
engineering~\citep{Wing}. Indeed, it has been precisely these concepts 
that have led to the ecosystem that has produced the LLM.

But ``computational thinking'' arose in the design of systems that had 
limited, carefully designed interaction with the outside world.  The real
world---the one that humans operate in---brings two major new sets of issues.
First, the real world is characterized by vast complexity and partial 
observability, such that coping with \emph{uncertainty} becomes a major issue.  
Secondly, there is a need to interact in social environments consisting
of \emph{strategic agents}.  I believe that these issues are of such vast importance 
that we cannot simply rely on existing computational design principles to address 
them.  Rather, we need to recognize that there are other ways to conceive 
of algorithm design.  I will propose two other thinking styles that complement 
computational thinking.  I refer to these styles---which are also the fruit 
of decades of experience---as ``inferential thinking'' and ``economic thinking.''  
The output of such thinking can be specific algorithms (and analyses of algorithms),
although the jargon is often different---in the inferential fields algorithms
are often referred to as ``procedures'' and in the economic fields algorithms
are often referred to as ``mechanisms.''  That these algorithms can be embodied
in computational devices is what has given them new power, but focusing solely
on ``computation'' misses the point.  It is the thinking behind the algorithms
that is important.

\begin{figure}[tb]
\begin{center}
\includegraphics[width=.45\textwidth]{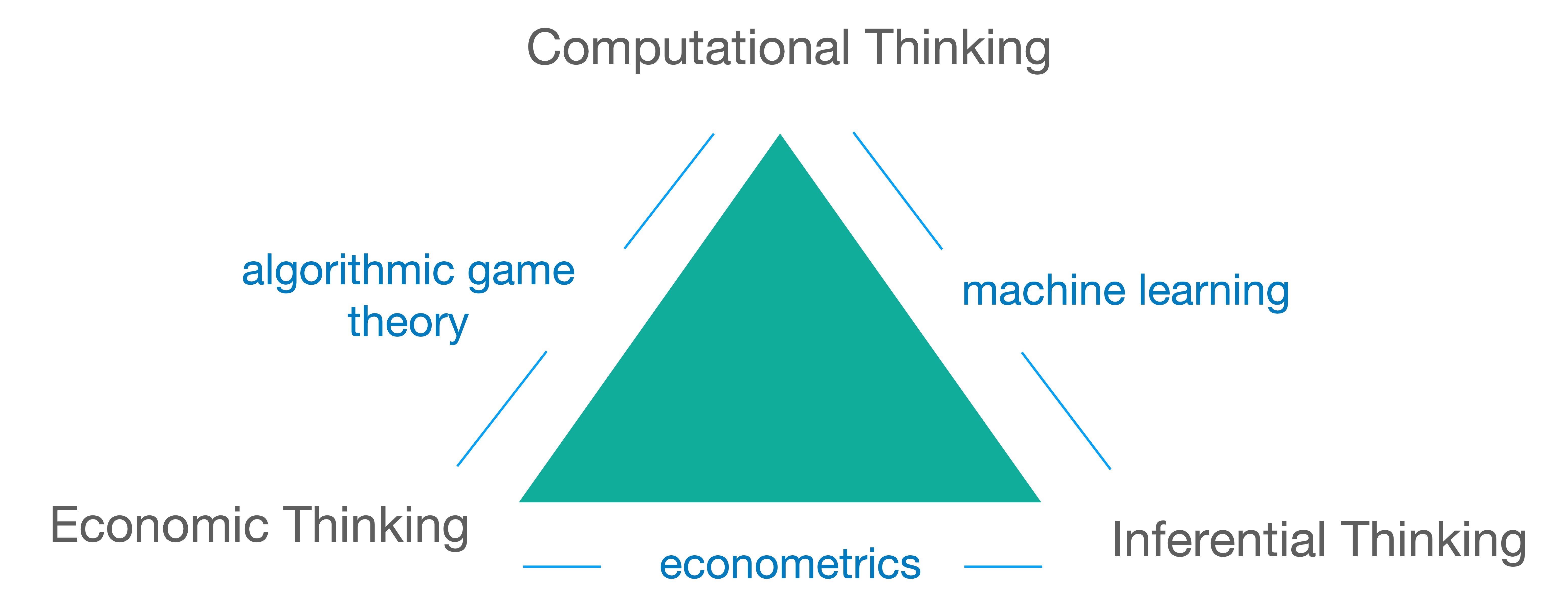}
\caption{Three core thinking styles that have come together in pairwise blends as academic disciplines.}
\label{fig:triangle}
%\Description{A diagram showing the tripartite blend of computational thinking, economic thinking, and inferential thinking.  The diagram also shows pairwise blends---machine learning, econometrics, and algorithmic game theory.}
\end{center}
\end{figure}
Before turning to concrete instances of design using these thinking styles,
let's return to the issue of coping with uncertainty.  Rather than attempting 
to delimit the notion of uncertainty with a precise definition, let's instead 
be informal and consider the broad problem of making decisions when some of 
the information that would be useful in making the decision is not available.  
Different fields have contributed different kinds of algorithms to cope with 
uncertainty.  Briefly, the field of statistics has focused on uncertainty 
arising from \emph{sampling}, whereby the data available for algorithmic 
processing is a subset of all of the data that would be useful in principle 
to solve some problem.  The field of economics has also made use of sampling 
ideas, but a distinctive focus of economics has been a different source of 
uncertainty---the \emph{information asymmetry} that arises when an agent 
interacts with another agent who possesses private knowledge and who may 
strategically reveal or conceal aspects of that knowledge in order to obtain
some desired outcome.  Information asymmetry does not tend to go away 
when the sample size grows.  Finally, another source of uncertainty 
comes from the ``when, where, and who'' of data collection.  When data is 
collected in the past, or at a distant location, or from a different source
than the current source, the uncertainty should grow.  The field of computer 
science has contributed tools to address this problem via the algorithmic 
concept of \emph{provenance}, whereby the origin and type of data are tracked 
systematically and coherently.  This gives substance to notions of \emph{relevance} 
in decision-making that are often treated informally in other fields.

It is useful to repeat this exercise with other properties that intelligent systems 
that operate in the real world might be hoped to exhibit, including robustness, 
coping with bias, understanding causality, and exercising an ability to do 
experiments.  The three thinking styles have contributed diverse and complementary 
perspectives in each of these cases.

In the next few sections I present vignettes that aim to capture the complementary 
nature of computational thinking, economic thinking, and inferential thinking.  
I specifically aim to demonstrate how blends of these thinking styles
yield powerful ways to approach system design for emerging data-catalyzed
systems.  As shown in Figure~\ref{fig:triangle}, pairwise blends of these 
thinking styles 
have already emerged as academic disciplines.  But each of the pairwise blends 
have made only limited use of the third ingredient, and accordingly they address
only part of the problem in systems involving people, machines, and data.  
What is needed is the tripartite blend.

\section{Computation and Inference in Database Design}
\label{sec:database-design}

Let us begin by introducing the perspective of ``inferential thinking'' in a 
stylized database problem.\footnote{See Figure~\ref{fig:database} in Appendix A 
for a figure that provides a visual accompaniment of the following discussion 
in the form of a set of flow diagrams.}  Consider a bank that maintains a database 
in which the rows correspond to clients and the columns correspond to financial 
data associated with each client.  The bank or others (e.g., auditors) may wish 
to perform various operations on the data, from simple calculations such as finding 
the average balance across clients, or more elaborate computations involving spotting 
unusual transactions.  The bank may also wish to provide a privacy guarantee to 
clients in response to such queries and thus may incorporate randomization.  
It will also be necessary to track provenance, provide interfaces for clients, 
and provide long-term storage of data.  In short, a great deal of computational 
thinking will need to go into the deployment of such a database.

What might we mean by ``inferential thinking'' in database design?  On the one hand, 
the database can implement statistical operations, such as computing a standard 
deviation or a linear regression.  Such computations are in the realm of what a 
statistician would call ``descriptive'' but they are not necessarily examples of 
``inferential thinking.''  To clarify the distinction, consider a different database 
in which the rows are patients in a hospital and the columns are vital signs for 
the patients, as well as indicators of treatment and responses to treatments.  
Now, at query time, we might like to ask how likely a particular patient is to 
respond favorably to a particular treatment.  In contrast to the banking example, 
we are probably not interested in patients who were in the original dataset---we're 
interested in ``new'' patients who come from the same population as the original 
patients.  Indeed, the original patients may be dead and gone, but the data is still 
valuable.  ``Inferential thinking'' refers to the design and analysis of algorithms 
that can extract this value.  It requires consideration of the underlying population, 
the set of possible queries, and the design of the sampling operator. It involves 
methods for checking whether the assumptions made in the design are reasonable post 
hoc.  Although the result of such design methodology is a set of algorithms, the 
thinking behind the design and analysis goes beyond ``computational thinking'' in 
its focus on entities that haven't been seen before.  For such entities the goal 
is not only to make a prediction, but also to provide a measure of uncertainty 
regarding that prediction.

More generally, inferential thinking involves characterizations of populations via 
a generative model, attempting to delineate underlying mechanisms by which data might 
arise, and choosing a model that fits the data well.  Such efforts often fall under 
the topic of \emph{causal inference}, where a key concept is the ``what if'' 
question---what if the database were different in some way from the data we 
collected?  What if a patient had been given the treatment rather than the control?  
Can estimates of population-level treatment effects be obtained from the sample 
data? These issues are subtle; see, e.g., \citet{HernanRobins}.

\section{Inference and Incentives}

Let us now bring economic thinking into the picture.  There are many existing 
connections of economics to computation; most notably, the field of algorithmic 
game theory~\citep{AGT}.  Moreover, within economics proper, the fields of mechanism 
design~\citep{Hurwicz} and information design~\citep{Bergemann} have been centered 
around algorithm analysis and design for several decades.  My goal is to highlight 
the opportunities that arise when the kinds of algorithmic ideas that one finds 
in these fields are juxtaposed with inferential and computational thinking, 
specifically in the design of large-scale, collectivist machine learning system.

Microeconomics focuses on the choices of strategic agents who are pursuing goals 
that may be entirely personal.  There is also a focus on the overall social 
welfare that can be achieved when self-interested agents interact.  Thus the 
issues that drive algorithm design in economics include information asymmetries, 
incentives, social goals, and solutions expressed as equilibria rather than optima.  

Let us begin by considering incentives, specifically in the context of inferential 
problems involving data analysis.  Such problems arise often in real-world ML 
deployments---even if they are not often not treated explicitly.  They arise in 
particular when the suppliers of data are agents who have strategic interests in 
the outcome of data analysis~\citep{PerdomoEtAl}.  This may lead to a misalignment 
between the goals of the agent and the goals of the data analyst, and lead to 
competition among multiple agents.  In such settings, the system designer will 
need to consider the design of incentives that will shape behavior; in particular, 
inducing agents to participate truthfully by sending actual data rather than 
falsified data or sending data that is selected in some strategic way.

The theory of incentives builds on game theory, which is a mathematical description 
of strategic behavior~\citep{Myerson}.  Indeed, the field of mechanism design, which 
encompasses the study of incentives, can be viewed as the inverse of game 
theory:  Whereas game theory aims to predict the outcome when strategic agents 
interact---with the outcome expressed as various kinds of equilibria 
(e.g., Nash equilibria for simultaneous play and Stackelberg equilibria for sequential 
play)---mechanism design starts with a desired outcome and asks what game would 
deliver that outcome as an equilibrium.

Sequential play is of particular interest for large-scale collectivist systems
given that agents are likely to act asynchronously in such systems.  Focusing on 
just two agents, one agent (referred to as a Leader) plays first, and the other 
agent (referred to as a Follower) plays next, with the Leader anticipating the 
Follower's response.  The uncertainty that is present in this situation is 
one that differs from statistical uncertainty.  Known as \emph{information asymmetry}, 
it reflects the fact that agents know different things and that there are 
strategic reasons to withhold one's knowledge in a transaction.  This kind 
of uncertainty does not go away by mere sampling; rather, it requires the design 
of an economic mechanism.

One mechanism that is appropriate for sequential play is known as a 
\emph{contract}~\citep{Laffont}.  Briefly, the idea is that the Leader 
does not simply play a single action (e.g., offer a price for some good), 
but rather presents a menu of options to the Follower which consists of 
a set of services and prices.  The Follower uses their private knowledge 
to pick the best option for themselves, and if the Leader has designed the 
menu well, then many Followers will find an appealing option in the menu. 
Compared to a mechanism in which a single fixed price is chosen, a contract
can deliver higher revenue given that some Followers may opt for a relatively
high price in exchange for the services being offered (those Followers with 
a high ``willingness-to-pay'').  Moreover, contracts can deliver high social
welfare (the aggregrate of the difference between willingness-to-pay and 
the actual amount paid).  Success in achieving such a criterion is expressed 
in terms of an equilibrium concept; namely, the Stackelberg equilibrium.  
Algorithm design involves specifying mechanisms (typically decentralized) 
that yield good Stackelberg equilibria.

Classical contract theory does not incorporate a role for inference from data, 
but an emerging field of ``statistical contract theory'' does precisely 
that~\citep{BatesEtAl}.  As an example, let us consider a sequential setting 
in which the Leader is a buyer who wants to perform hypothesis testing---making 
``buy'' and ``no buy'' decisions for a sequence of products, where the products 
are supplied by self-interested suppliers (in the role of Followers).
The buyer may be viewed as a marketplace and the testing is needed to determine
which products go to market.  Suppose that some of the products are of high 
quality and others are of low quality.  The buyer does not know which products 
are of high quality and may therefore collect a certain amount of data (for 
example, using a focus group) to make their decision.  In doing so,
they will be be making statistical errors, given limitations on the amount of 
data collected (which may be costly).  Specifically, there will be false positives 
and false negatives.  We imagine that the buyer's goal is to minimize some 
function of these errors.  To do so, they have to cope with information 
asymmetry.  The suppliers may know which of their proffered products are 
of high quality and which are of low quality or may have some rough prior 
information---in particular they may have invested more effort in making 
some of the products---but they are not incentivized to reveal this information 
to the buyer.  Indeed, their hope is that some of their low-quality products 
may end as false positives, which is profit for them.  

Statistical contract theory aims to design contracts that incentivize the
suppliers to send in products that are more likely to be of high quality,
such that the overall mix of products has a controlled statistical error.
The contract is a menu of options, where each option involves various costs
(e.g., shouldering some of the data-collection burden) and various licensing
terms (e.g., responsibilities vis-a-vis customers).  Some options will be
more or less risky and more or less lucrative.  The supplier uses their
internal knowledge to make the choice, essentially making a bet on the 
possible outcomes.  The buyer sets up the contract so that it is 
\emph{incentive compatible}---in particular, if an item is actually of low 
quality, the expected total profit for the supplier is nonpositive, so
the system cannot be gamed.

\citet{BatesEtAl} prove that statistical contracts are incentive-compatible
in this hypothesis-testing problem if and only the options can be expressed as 
\emph{e-values}.  An e-value is a function of data that is less than or equal 
to one in expectation if a null hypothesis is true~\citep{RamdasWang}.\footnote{E-values 
can viewed as an alternative to p-values, which are tail probabilities under 
the null hypothesis rather than expectations.} They have a betting interpretation 
as the multiplicative factor by which wealth increases (in expectation) under 
the null hypothesis.  More generally, when data arrives sequentially in time, 
the appropriate function of data is the nonnegative supermartingale, which is 
an e-value at any stopping time and which can be viewed as the accumulation 
of evidence over time.   What the result of \citet{BatesEtAl} shows is that 
an important inferential concept (e-values for hypothesis testing) is closely 
linked to an important economic concept (information asymmetry in contract design).

\section{Multi-Way and Multi-Layered Markets, the Internet, and Foundation Models}
%\label{sec:markets}

I now turn to a consideration of examples in which the three thinking styles come 
together in system design.  Rather than attempting to provide formal treatments 
of such blends,  I focus on qualitative examples that have real-world significance. 

The Internet is a huge collection of text, images, video, and links and, as such,
can be viewed as a source for data analysis.  But it is also a place where 
interactions among humans occur, where creative collective activity such as 
Wikipedia has taken place, and where markets have arisen.  Many of these 
markets have created real social value, but many are also defective along 
one or more dimensions---in particular, in their inability to reward creators, 
to value data as an economic good, to create trust, and to disincentivize socially 
harmful behavior.  Part of the problem is that little thought has been given to 
mechanism design in building out the Internet.  An exception is advertising markets, 
which have created revenue but which have been a mixed bag with respect to social 
welfare.  In this section, I will discuss other markets that exist on the Internet,
or could exist, from the point of view of our three thinking styles, highlighting
opportunities for improved social welfare.

\subsection{Recommendation systems}
\label{sec:rec-systems}

Recommendation systems are a classical example of ML systems that are collectivist.  
In one instantiation of a recommendation system, one considers a graph which links 
customers on one side and products on the other.  Purchases are represented by edges 
between a customer and the products that they purchase, and graph-theoretic methods 
exploit similarity patterns in the graph to make targeted recommendations.

Although recommendation systems do bring ML closer to microeconomic considerations,
they are limited as microeconomic entities---in particular, no money changes hands.
Good recommendations may lead customers to make purchases, but conceptually this is
just a way to make an existing market for physical goods more efficient.  There is
no strong need for consideration of incentives.

\begin{figure}[tb]
\begin{center}
\includegraphics[width=.37\textwidth]{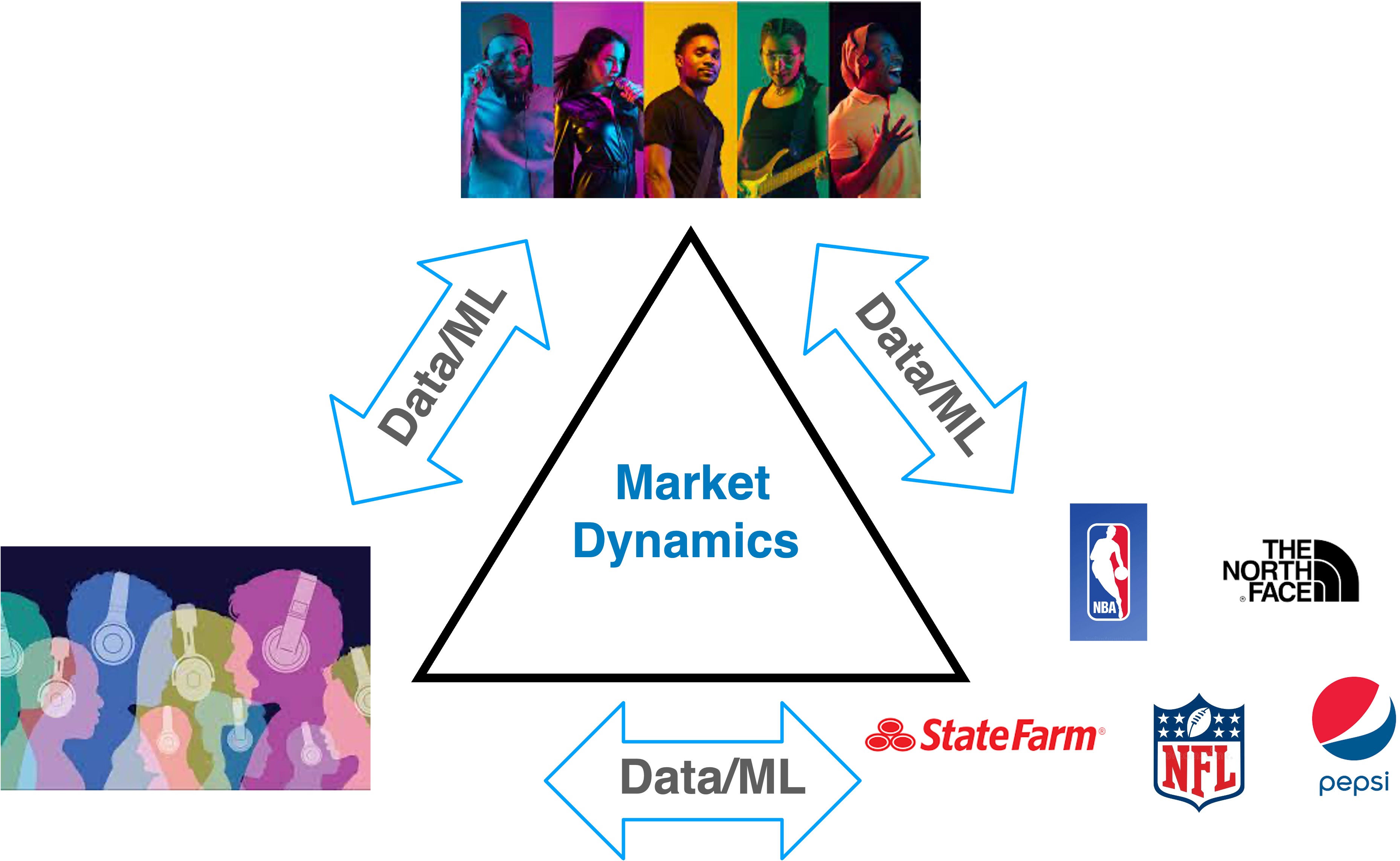}
\caption{A depiction of a three-way market for recorded music which links musicians, 
listeners and brands.}
\label{fig:united-masters}
%\Description{A depiction of a three-way market for recorded music which links musicians, listeners and brands.}
\end{center}
\end{figure}

Let us consider a market that is currently in the midst of technology-driven 
change---that of recorded music.  In the days of yore recorded music was a physical 
good but it has become a virtual good.  For virtual goods, the lack of economic 
mechanism design in recommendation systems is problematic, leading to an impoverished 
reward system for creators, who wield little market power.  Let us consider an 
alternative.  In Figure~\ref{fig:united-masters}, I depict the design of a 
three-way market for recorded music.\footnote{This market is the architecture 
underlying the company United Masters, where the author is a board member and 
consulted on the market design.} At one vertex are musicians, who supply songs, 
and at a second vertex are listeners.  The musicians and listeners are linked 
by a classical ML-based recommendation system.  Critically, there is a third 
vertex, which are brands.  Brands often make use of music in products and outreach 
and they need well-chosen music which fits their image and connects well with 
the demographic they cater to.  Additional recommendation systems, also powered 
by trained ML models, provide these connections.  Moreover, critically, the 
overall design incorporates incentives.  When a brand needs a song, they are 
supplied with a song from a particular artist (using an ML model), and the 
artist is paid in that moment.  Audience reaction is measured.  Other brands 
can see that reaction, and if it happens to be aligned with a demographic 
they're also interested in, then they are incentized to reach out to the artist 
and partner with them.\footnote{It is worth noting that United Masters has 
signed over 1.5 million musicians to date, and their music is used by brands 
such as the NBA, Bose, and State Farm.  This is an instance of a collectivist 
AI system that has created jobs.}

Note the difference with the classic online business model for recorded music, where 
musicians upload their music to the cloud and it is streamed to listeners for free.  
Money is made by the platform, via subscriptions or by advertising, but there is no 
direct connection between producer and consumer, and there is a weak incentive for 
the platform to send money back to the musicians.\footnote{Indeed, there's a strong 
incentive for the platform to use generative AI tools to replace the musicians!}

Different ways of conceiving of market dynamics for a learning-based online service 
can have rather different outcomes in terms of social welfare.

\subsection{Data markets}

\begin{figure}[tb]
\begin{center}
\includegraphics[width=.45\textwidth]{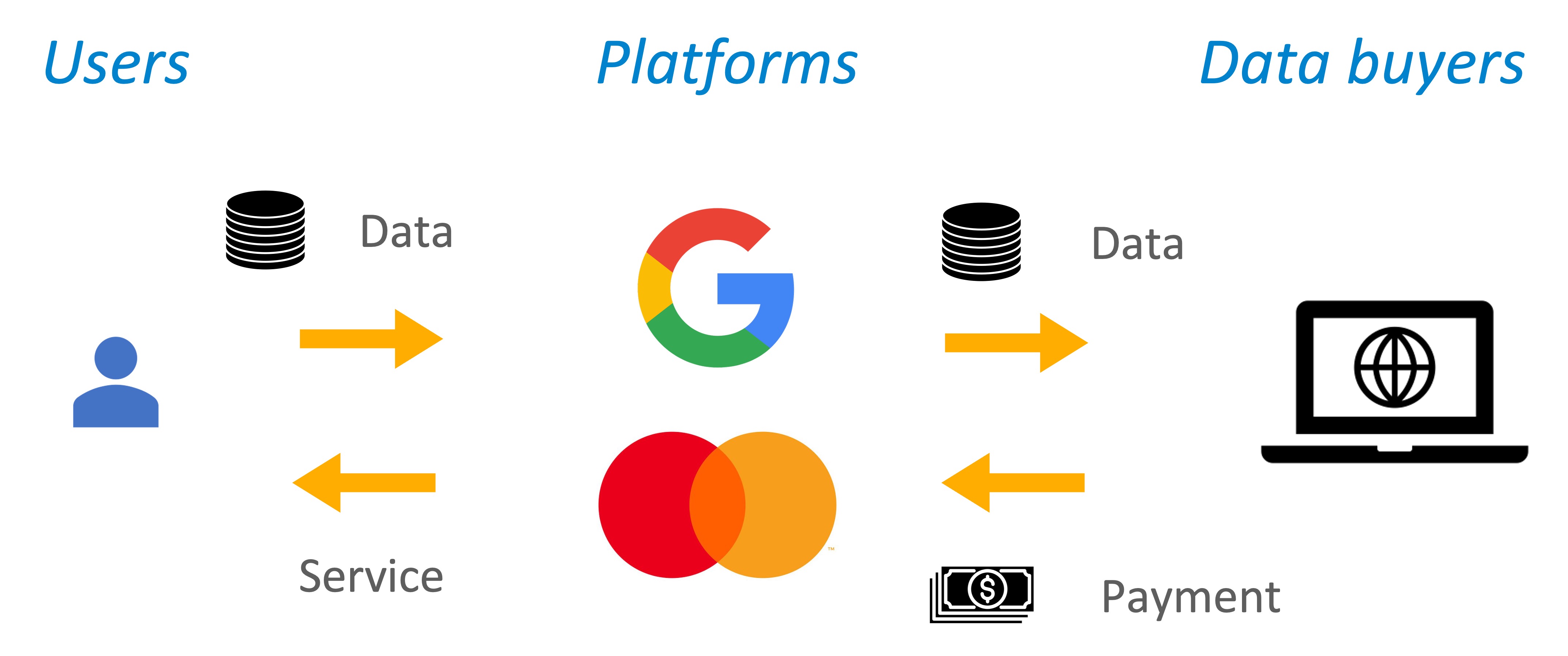}
\caption{A three-layer market in which a platform provides services,
and also sells data to third-party buyers.}
\label{fig:three-layer}
%\Description{A depiction of a three-layer market for which links a user, a set of platforms that provide services to the user, and third-party buyers who pay for data from the platforms.}
\end{center}
\end{figure}

Let us now consider a different problem domain that features a three-component market, 
in this case taking the form of a layered structure~\citep{FallahEtAl}.  
In Figure~\ref{fig:three-layer}, a user is shown interacting with a platform, 
which provides a service and receives a payment in return (concretely, let us 
suppose that the platforms provide access to credit for a fee).  We imagine that 
the platform can learn from the data that it obtains from the user and thereby 
improve the service.

Thus far we have a market where data plays an informational role, but data is not 
a transacted good.  Both the user and the platform are incentivized to engage 
in this market, depending on the details of the service and the fee.  But in many 
such situations, the platform does not make enough revenue from the fees it 
collects to realize a profit~\citep{Cross}.  Thus, it turns to a set of third-party 
data buyers.  These buyers wish to acquire data for their own purposes (such as 
carrying out market research).  The platform acts as a supplier, and the data 
becomes a transacted good.  It will be priced according to its value to the data 
buyer and other factors.  Again, the incentives are aligned for the platform and the buyer.

But now consider the overall system and ask whether the user is incentivized to 
participate in this layered market.  A new issue has arisen---the user has lost 
control over their privacy in this market.  Whereas before the platform could 
be held to a contract so that the user has a guarantee that their private data 
is being used in a limited way---and they are receiving a service and therefore 
are presumably willing to incur some privacy loss---now the user is told that 
third-party data buyers are acquiring his or her data and all bets are off.  
No new service accrues to the user in exchange.  The privacy loss can be 
unbounded and substantial.

Thus, the user is likely to walk away, and the design needs to be elaborated 
for the market to function.  Let us imagine, for example, that platforms decide 
to provide a formal guarantee of privacy when sending data to third-party data 
buyers.  Concretely, this would mean that noise is added to the data of a 
magnitude that is contractually specified (and can be audited).  Although the 
noise level could be subject to government regulation, let's instead leave the 
choice in the hands of the platforms.  A platform is incentivized to provide 
a nontrivial level of noise, because users will shop among platforms to find 
one that provides a desirable level of privacy in conjunction with an effective 
service.  Moreover, such a platform, by accruing more users, will collect 
more data and can make further improvements to its service.  On the other hand, 
data buyers are averse to noise and will presumably pay less for data from 
the platforms that provide a stronger privacy guarantee.  Thus, there is a 
conflict for the platforms.  The way to understand how the conflict will play 
out is to model the overall system as a game (it is a generalized Stackelberg 
game) and find its equilibria.  This requires specifying utility functions 
or preferences for the various players.

In this scenario, both the platforms and the data buyers will likely be ML 
systems, learning from the data that they receive.  Moreover, data is an 
endogeneous part of an overall system in which learning, data, and human 
preferences all interact.  Understanding how the overall system will 
behave---and in particular whether it will work at all---requires a 
collectivist perspective that combines ML with economics.

\subsection{Foundation models, bias, and local knowledge}
\label{sec:bias}

Foundation models are large-scale ML systems that aim to make high-quality
predictions in domains of major scientific or societal interest.  For example,
LLMs are foundation models for natural language and AlphaFold is a foundation 
model for protein structure~\citep{AlphaFold}.

The phrase ``high-quality predictions'' warrants some discussion.  Let us 
consider first scientific domains, where there is in principle a ground truth
to compare to.  For example, AlphaFold exhibits high overall accuracy when 
compared to protein structures that have been determined in the lab.  This 
does not mean, however, that it is uniformly accurate.  Rather, it is accurate
on training data and test data that have been obtained from past scientific
investigation.  Unfortunately, scientists are often interested in phenomena
that are on the edge of knowledge, such that there may be little past data to 
support accurate prediction.  Indeed, \citet{PPI} showed that AlphaFold can
given highly biased confidence intervals (intervals that are overly narrow 
and do not cover the ground truth) for certain queries involving proteins
that exhibit quantum fluctuations (where there are not many ground-truth
measurements).  \citet{PPI} also demonstrated that such biased confidence
intervals arise in a wide range of other scientific domains.

This problem can be addressed via a technique developed by \citet{PPI} that
is known as \emph{prediction-powered inference}.  This is an inferential
algorithm in which assessments of uncertainty obtained from global foundation 
models are adjusted based on local ground-truth measurements that are possessed 
by a local agent.  Such local knowledge may involve measurements that were 
not available in the training of a foundation model or may simply reflect a 
particular (desirable) bias of the local agent.  PPI-corrected confidence
intervals provably cover ground-truth estimands, under standard statistical
assumptions.

More generally, in domains involving strategic interactions between humans
and non-humans, an agent that queries another agent (which may be a 
foundation model) will generally need to be concerned about bias.  Moreover, 
the strategic nature of the interaction means that the bias may have been 
created willfully, to align with the goals of the other agent.  In this
context, PPI can be viewed as something more than a debiasing technique.  
If the agent providing the data or the foundation model is aware that 
the receiving agent will be making use of local ground-truth data, then 
that agent will be disincentivized from providing data that is significantly 
biased. The agent will also be incentivized to expand the scope of their 
data or their model to meet the receiving agent's needs (if they want to 
continue to interact with that agent).

Still more generally, ambiguity in what constitutes the ``correct response''
will arise from the fact that data and knowledge are often local, contextual, 
and fleeting.  Solving one's local problem will often involve a blend of 
information from outside sources with information that is only available locally
in space and in time.

\section{Discussion}

The study of multiple agents in computer science is by no means new, and 
researchers in fields such as multi-agent ML, human-computer interaction, 
and algorithmic game theory will recognize that it is their work that is 
being promoted here, as are the perspectives of the social sciences broadly 
speaking.  Moreover, there are antecedents of our arguments in the study 
of ``collective intelligence''~\citep{TumerWolpert,Parkes,Malone}. That 
literature has two main branches, one focused on qualitative, experimental 
work with groups, and the other focused on artificial collectives with 
designed agent utilities. My focus is different. I want to uncover algorithmic
design principles for emerging real-world ML-based systems in which many 
of the participants are human and many are non-human. The goals and utilities 
of the humans are to be understood and respected, not designed. The principles 
should be simultaneously computational, economic, and inferential.

Although ML-powered markets can be expected to differ in important ways from 
classical markets, it's worthwhile to recall some of the appealing features of 
classical markets.  First, they provide a form of uncertainty reduction---buyers 
can count on products to be available and plan accordingly.  Second, they cope 
with heterogeneity.  Third, they create new roles on an as-needed basis.  
It is early days for ML-powered markets, but it is not hard to imagine many
new roles that can be expected to arise, akin to those that have arisen in 
previous eras of technology but based around data and learning---auditors, 
brokers, aggregators, sellers, buyers, artists, forecasters, insurers, and 
explorers.  These roles will give rise to personalized services, economies 
of scale, and appropriate touch points at which regulatory control can be 
exerted to mitigate those biases that are proscribed by legal or ethical 
considerations.

Design principles based on our tripartite blend make it easier to discuss issues
of interest to both individuals and collectives, such as privacy, fairness, 
ownership, alignment, reputation, and transparency.  The blend allows these 
issues to be treated as tradeoffs rather than being reduced to black-and-white 
distinctions.  For example, differential privacy is a computational method 
that adds noise to data to guarantee privacy according to a particular 
definition~\citep{DworkRoth}.  But this is only part of the story.  An 
individual's decision to ask for a certain level of differentiable privacy 
in an interaction will trade off with the other costs and benefits that the 
individual expects from the interaction.  Moreover, inferential issues arise 
whenever noise is being added to data, and there are quantitative tradeoffs 
that can be formulated that link inferential accuracy and privacy~\citep{DuchiEtAl}.  
Making full use of the tripartite blend will generally involve solutions that 
take the form of tradeoffs.

Finally, for AI to grow into a mature engineering discipline that delivers systems 
that yield value to humans in real-world settings, it will need far more than blends 
of existing algorithms (and far more than just more data and more compute).  In this 
regard, it's useful to learn from the history of the fields of chemical engineering 
and electrical engineering, which brought the complex phenomena of chemical reactions 
and electromagnetism under control.  This was achieved by developing modular, transparent 
design concepts that were appropriate for the phenomena.  The modularity allowed large 
systems to be designed piecemeal, allowed system failures to be diagnosed and repaired, 
and allowed multiple stakeholders to participate in the evolution and regulation of 
systems.  We are far from such design concepts in the current stage of development of 
AI.  Moreover, chemical engineering and electrical engineering had at their foundations 
Schr\"odinger's equation and Maxwell's equations, solid foundations that could guide 
the development of simplifying modular approximations in the face of exceedingly complex 
phenomena.

For AI, we certainly have exceedingly complex phenomena---cognitive, social, commercial, 
and scientific---but we do not have the equivalent of Maxwell's equations as a guide.  
We are winging it.  Going forward, we therefore need the very best of our overarching, 
hard-won general scientific and humanistic principles---including rationality, experimentation, 
dialog, openness, cooperation, skepticism, empathy, and humility---as daily companions 
on the journey ahead.

\section{Acknowledgments}

I would like to acknowledge helpful discussions with Anastasios Angelopoulos, Francis Bach, Stephen Bates, David Blei, Alireza Fallah, Nika Haghtalab, Guido Imbens, Meena Jagadeesan, Barbara Rosario, Ion Stoica, Steve Stoute, Hal Varian, Rakesh Vohra, Serena Wang, and Tijana Zrnic. This work was funded by the European Union, ERC-2022-SYG-OCEAN-101071601. Views and opinions expressed are however those of the author(s) only and do not necessarily reflect those of the European Union or the European Research Council Executive Agency. Neither the European Union nor the granting authority can be held responsible for them.
I also wish to acknowledge funding by the Chair ``Markets and Learning,'' supported by Air Liquide, BNP PARIBAS ASSET MANAGEMENT Europe, EDF, Orange and SNCF, sponsors of the Inria Foundation.

\bibliographystyle{ACM-Reference-Format}
\bibliography{collectivist-ai}

\appendix

\section{Inference and Privacy in Databases}
\begin{figure}[tb]
\begin{center}
\includegraphics[width=.9\textwidth]{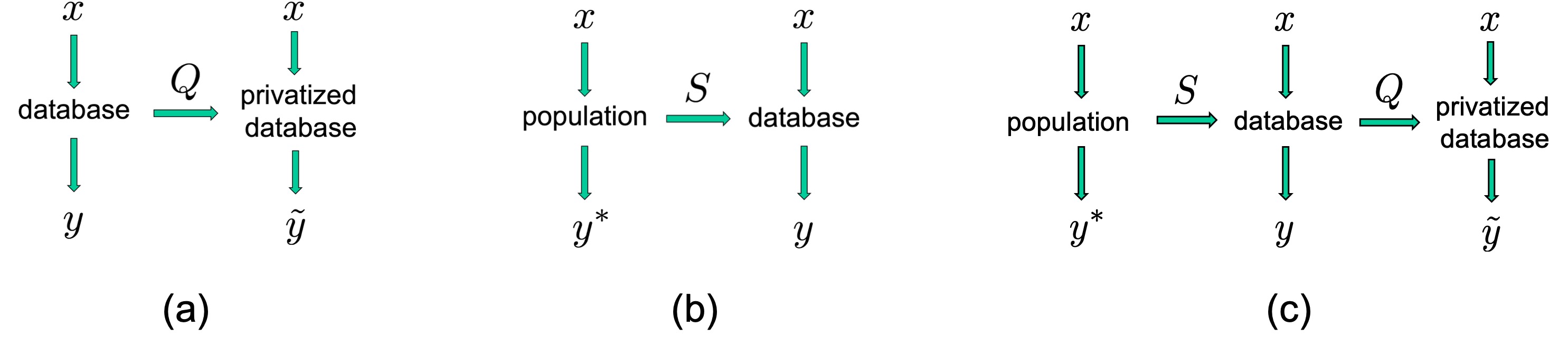}
\end{center}
\caption{Database design problems that highlight the complementary nature of computational thinking and inferential thinking.}
%\Description{Three database design problems presented as flow diagrams, and incorporating both privacy constraints and sampling from a population.}
\label{fig:database}
\end{figure}

We discussed three database design scenarios in Section~\ref{sec:database-design}.
Figure~\ref{fig:database} provides a visual depiction of these scenarios.

In Figure~\ref{fig:database}(a), we depict the standard privacy-preservation problem, 
in which the problem is to develop a (randomized) algorithm $Q$ that provides privacy 
guarantees while ensuring that the privatized response $\tilde{y}$ to a query $x$ 
is not too far from the ``true response'' $y$.  The latter refers to the response 
that would be obtained without privacy constraints.

Note that the diagram is not meant to suggest that the randomized algorithm $Q$, 
which concretely could be a differentially private operator, needs to involve 
transforming the entire database.  Rather, using standard ideas from the differential privacy 
and online learning literatures~\citep[cf.][]{DworkRoth}, we envision $Q$ as being 
called on an as-needed, possibly query-dependent basis.

Figure~\ref{fig:database}(b) presents an inferential perspective, in which the database 
is assumed to arise from an underlying population, under a (randomized) sampling operator 
$S$.  The sampling operator may be given or designed (or it may be some blend).  The 
goal is to ensure that $y$ is not too far from the ``true response'' $y^*$.  Here the 
latter is an unobservable population quantity.  Statistical theory is needed to make 
this kind of guarantee.

Of course, in real-world problems there is often a need to combine these two thinking styles.  
A more general setup is one in which the goal is predictive inference for a new patient, 
but where we also want to protect the privacy of the individuals in the original dataset, 
perhaps because their offspring will have an interest in ensuring their own privacy.
Figure~\ref{fig:database}(c) presents the privacy-preserving-inference problem, where we 
wish to make predictions for new individuals who are sampled from the underlying population 
while providing a privacy guarantee for the individuals who were in the original database. 
Here, we want $\tilde{y}$ to be close to $y^*$~\citep[see, e.g.,][]{BlumEtAl,DuchiEtAl}.

Privacy appears in several of our vignettes, including the database example and the
three-layer data market.  This is not surprising, given its prominence in public 
discussion of information technology, including in legal circles, and given that
differential privacy and other computational procedures have provided ways to quantify
certain aspects of privacy.  As we've noted, this opens up a discussion of tradeoffs
involving economic value and inferential value.  

It's also worth pointing to another computational framework for treating issues
surrounding privacy that is known as \emph{contextual integrity}~\citep{Nissenbaum}.
Contextual integrity aims to provide formal reasoning strategies for evaluating
data flows in social contexts.  This is a different direction than the quantitative 
and algorithmic computation-economics-inference focus of the current paper, but the
goals of connecting social desiderata to technological design are shared, as is the
emphasis on context.  It is of significant interest to engage in a deeper comparison
of the current perspective with that of contextual integrity, not only for privacy
but for related economic and personal desiderata in AI systems.

Finally, there is much to discuss regarding the interactions of privacy with incentives
and trust.  This is a domain in which hardware and cryptography can provide some 
assistance---if an individual is assured that his or her data is being processed 
using some form of trusted execution environment and/or secure multi-party
communication algorithm~\citep{Liao}, then that individual can be viewed as 
incentivized to tolerate additional privacy risk.

\section{The Missing Middle Kingdom in AI Education}

What are the implications of the tripartite blend for undergraduate and
graduate education?  A naive response would be that students should simply 
take a collection of classes in computer science, economics, and statistics.  
But this response is lacking in several ways.  First, these three disciplines
have not cornered the market on ``computational thinking,'' ``economic thinking'' 
and ``inferential thinking.''  Important instances of these thinking styles
can be found throughout academia.  Second, classes in computer science, 
economics, and statistics  often aim to solve problems from previous eras, 
and the tripartite blend aims at emerging problems.  Third, by simply asking 
students to take more classes we are putting the onus on the students to 
put together the concepts they are learning in the context of real-world 
problems.  Educators should be shouldering more of the burden of innovative 
thinking in an era of major change.

It is noteworthy that the phrase ``computational thinking'' arose in part 
to support an expanded presence of computational ideas in academic 
curricula~\citep{Wing}.  In 2015 my colleagues and I at UC Berkeley 
built on that perspective to design a new course, Data 8, that aimed to 
introduce college freshmen to a blending of ``computational thinking'' 
and ``inferential thinking.''  To give a concrete example of what this
blend can mean at the freshman level, consider a problem in which the goal
is decide whether a treatment is effective based on a comparison of two
groups of outcomes.  Students were presented with two columns of numbers, 
labeled ``treatment'' and ``control,'' and asked whether it seemed 
plausible these columns were significantly different (i.e., there was 
an effect of the treatment), or not.  To aid this assessment, students 
first learned how to make histograms (learning just enough Python 
programming to carry out this task).  Comparing the histograms visually 
gave a sense of possible differences.  To take a step further and 
address an inferential question, we introduced the \emph{permutation 
test}.  This test involves some ``what-if'' thinking, where one 
considers the possibility that the columns of numbers come from the 
same source.  In that world, one can lump the two columns together, 
permute the resulting single column, split it into two new columns, 
and compute whatever comparative statistic between the columns that 
one likes.  (We used the variation distance, which is a simple comparison
between histograms).  Doing this repeatedly yields a sampling distribution,
which can be plotted as a histogram.  Placing the statistic of the original
two columns on this histogram, one can assess whether this statistic seems
likely given the hypothetical ``what-if'' thinking that the simulation process 
embodies.  Finding that statistic out in the tail of the histogram suggests 
that that hypothetical assumption may be false.

Clearly there is a lot of inferential reasoning going on in this example,
and it can be blended with computational concepts in ways that go beyond just 
plotting histograms.  For example, how does one permute a column of numbers in 
Python?  There is a naive $O(n^2)$ algorithm to do it, but there is also 
a more sophisticated $O(n)$ algorithm.  Aha!  Then, critically, with this 
blend in hand, we returned to the students and asked them to find their 
own problems, and accompanying datasets, where ``treatment'' and ``control'' 
were whatever real-world distinction they want to investigate.  (E.g., 
does water quality vary between more wealthy parts of a city and less 
wealthy parts? or Was there more deforestation in the Amazon this past
year as compared to five years ago?).  Such real-world investigation 
was exciting for a diverse set of students, in ways that the problems 
studied in classical computer science curricula are sometimes not 
(e.g., game-playing programs).  Data 8 became the fastest-growing class
in the history of UC Berkeley, and is currently taken by over 1500 students
each semester.

Returning to the original design process for Data 8, it is natural to
consider adding ``economic thinking'' to the mix.  As I have emphasized 
in this article, in many problems data arises out of interactions among 
humans, and the algorithms that analyze such data should take such
interactions into account.  One can well imagine that basic ideas such
as matching markets, contract design, and auctions could be integrated
into the algorithms-for-real-world-problem-solving perspective of Data 8.

Such material would retain the core focus on algorithms, but it would
also open up coverage of social issues (such as privacy, fairness, ownership, 
alignment, reputation, and transparency).  It would also demonstrate that
the world of algorithm design---the key leverage point that we have on 
the evolution of modern technology---is worth understanding by a broad 
swath of students, not just computer engineers.  Understanding the role
of algorithms that address economic and inferential issues can help engender
a dialog that is more multi-directional than simply teaching computational 
concepts to constituencies outside of computer science.

A related virtue of the tripartite blend is that it is inclusive of a wide
range of disciplines that have already developed connections to computer 
science, economics, and statistics.  Indeed, from these three starting
points, one arrives quickly at disciplines such as cognitive science, 
social psychology, law, sociology, public policy, mathematics, physics,
biology, and the humanities.  These are all disciplines that can provide
additional perspectives that are essential in shaping human-centric technology.  
To take but one example where these perspectives link to our discussion, 
consider behavioral economics, which links cognitive and social psychology 
with economics, aiming to overcome a perceived overreliance on unrealistic 
assumptions of rationality in classical economics. Here too, however, 
existing concepts need to be revisited in the data-centric era.  Recalling 
that humans provide the data on which AI artifacts such as LLMs are based, 
we see that behavioral data is in some sense already baked into learning-based 
systems.  Blending the massive-but-messy information in such data with the 
more traditional, experiment-driven insights coming from behavioral economics 
is an important agenda item for future research.

It is often suggested that there should be stronger links between computer
science and the humanities.  But in practice this often boils down to having
computer science students take an ethics class or another elective, and to
having humanities students acquire programming skills.  The gap between computer 
science and the humanities is large, and it is difficult to design direct bridges.
This is due in part to the absence of connections that flow through the social
sciences in attempts to design such bridges.  Indeed, some social sciences have 
natural connections to computer science (e.g., economics), others have natural 
connections to the humanities (e.g., history), and in general the focus on humans 
and groups of humans provides essential perspectives.  Our tripartite blend can 
be viewed as providing a new kind of engineering core that aims to catalyze 
existing connections and spawn others.  It provides a middle kingdom between 
engineering and the humanities that can help refresh academic debates and help 
design forward-looking, inclusive curricula.

\section{LLMs, Uncertainty, and Collectives}

I have argued that the ability to cope with uncertainty---at both the individual
level and the collective level---is a hallmark of intelligence.  Recall that we
are considering a broad, informal notion of uncertainty that involves sampling, 
information asymmetry, provenance, and other ways in which information that is 
relevant to a decision may be biased, partially available, or stale.

Current-generation LLMs make little use of any explicit notion of uncertainty.
They can give the impression that they are quantifying uncertainty and reasoning 
inductively, but this is in part an illusion, arising from the fact that the humans 
who produced the data on which the learning algorithms are based employed terminology 
and arguments associated with reasoning under uncertainty.  It is far from clear 
that LLMs have extracted useful general principles for managing uncertainty from
such data, particularly given that individual humans do not necessarily follow
normative rules~\citep[see, e.g.,][]{Kahneman, Griffiths}.  Moreover, some of the
known problems with LLMs, such as systematic overconfidence~\citep{Sun}, are not
exhibited by individual humans to the degree that they are exhibited by LLMs.  
It remains an important item for the research agenda to develop a deeper understanding 
of the cognitive science of uncertainty and the appropriate decision-theoretic
underpinnings for LLMs.

But, as I have argued, a cognitivist perspective is not sufficient for understanding
and coping with real-world uncertainty.  Moreover, I have suggested that LLMs can be 
usefully viewed as a collectivist entity rather than a singular intelligence.  In this 
section I provide further discussion of these two issues and their interrelationships.

A first point is that collectives play an essential role in changing the world in 
ways that shape and mitigate uncertainty. An individual human foraging for fruit 
may or may not succeed in finding fruit on any given day, but once a collective 
creates a market for produce, the uncertainty regarding finding fruit drops 
significantly.  A human can then depend on fruit being available and build on 
that certainty, perhaps opening a pastry shop.

Collectives also help to define which uncertainties an individual should focus on 
and how uncertainties impact the choice of actions.  Consider an example from behavioral 
psychology---a two-choice maze experiment in which a mouse learns that there is more 
food on the left branch than on the right branch in a ratio of 2/1.  The mouse now 
needs to decide which branch to take on the next trial.  A mouse that is 
decision-theoretically optimal would visit the left branch with probability one, 
maximizing its chances of finding food.  Real mice (and humans in analogous situations) 
often do something different, known as ``probability matching''---they visit the left 
branch twice as often as the right branch.

Why should that be?  Consider a setting in which there are many mice.  Here, if each 
mouse goes left with probability one, then the food on the right is an unexploited resource.  
A probability-matching algorithm rectifies this---each mouse independently chooses to 
go left versus right with a ratio of 2/1 and the result is high overall social welfare.  
An appealing explanation for its emergence is that it is an equilibrium that is found 
via evolutionary forces~\citep{Lo}.

In general, collectives are the appropriate level of analysis for many problems 
involving scarcity and uncertainty.  The solutions to such problems may be baked 
in by evolution, but more often than not they require communication and interaction 
among agents.  This leads us to microeconomic concepts but also requires us to
augment those concepts with a fuller toolbox for the management of uncertainty
(for example, the distribution-free methods such as conformal inference that have
been developed in recent years~\citep{Conformal}).  Thus we should envisage new 
kinds of markets in which data flows and distributed statistical analysis play 
key roles in shaping agent behavior.  Machine learning concepts will be crucial 
given the kind of dynamic, highly distributed systems that are our likely future, 
in which data will be increasingly partial, noisy, biased, local, contextual, strategic, 
and generally incoherent.  From this perspective, intelligence is not just about 
knowing things, and giving answers that are ``correct,'' but knowing how to act 
when one's knowledge is partial, and knowing how to interact with other individuals 
whose knowledge is partial.

In summary, the two themes of ``social'' and ``uncertainty'' are closely 
related.\footnote{It's worth noting that this is a point that has been 
emphasized in the field of finance~\citep{Merton}.} On the one hand, social 
environments create various kinds of uncertainty, including information asymmetry 
(some actors are better informed than others, in ways that depend on the problem 
at hand) and strategic obfuscation (am I being told the truth and how would 
I know?).  On the other hand, social environments permit cooperation and the 
sharing of information, mitigating uncertainty for everyone, and thereby improving 
decision-making and enabling long-term planning.  Real-world intelligence is 
as much a social, communications, economic, and cultural concept as a cognitive 
concept.

\end{document}